\documentclass[
    aps,pra,
    amsmath,amssymb,
    reprint,
    superscriptaddress
   ]{revtex4-1}
   \newcommand{\bv}[1]{
     \ensuremath{\boldsymbol{#1}}
   }

   \newcommand{\poly}{\mathrm{poly}}
   
    \usepackage{graphicx}
    \usepackage{color}
    \usepackage{dcolumn}
    \usepackage{siunitx}
    \usepackage{bm}
    \usepackage{braket}
    \begin{document}
    \newtheorem{theorem}{Theorem}
    \newtheorem{fact}{Fact}
    \newtheorem{corollary}{Corollary}[theorem]
    \newtheorem{definition}{Definition}
    \newtheorem{assumption}{Assumption}


    \title[]{Quantum analog-digital conversion}

    \author{Kosuke Mitarai}
    \email{mitarai@qc.ee.es.osaka-u.ac.jp}
    \affiliation{Graduate School of Engineering Science, Osaka University, 1-3 Machikaneyama, Toyonaka, Osaka 560-8531, Japan.}
    \author{Masahiro Kitagawa}
    \affiliation{Graduate School of Engineering Science, Osaka University, 1-3 Machikaneyama, Toyonaka, Osaka 560-8531, Japan.}
    \affiliation{Quantum Information and Quantum Biology Division, Institute for Open and Transdisciplinary Research Initiatives, Osaka University, 1-3 Machikaneyama, Toyonaka, Osaka 560-8531, Japan.}
    \author{Keisuke Fujii}
    \email{fujii.keisuke.2s@kyoto-u.ac.jp}
    \affiliation{Graduate School of Science, Kyoto University, Yoshida-Ushinomiya-cho, Sakyo-ku, Kyoto 606-8302, Japan.}
    \affiliation{JST, PRESTO, 4-1-8 Honcho, Kawaguchi, Saitama 332-0012, Japan.}

    \date{\today}

    \begin{abstract}
        Many quantum algorithms, such as Harrow-Hassidim-Lloyd (HHL) algorithm, depend on oracles that efficiently encode classical data into a quantum state.
        The encoding of the data can be categorized into two types; analog-encoding where the data are stored as amplitudes of a state, and digital-encoding where they are stored as qubit-strings.
        The former has been utilized to process classical data in an exponentially large space of a quantum system, whereas the latter is required to perform arithmetics on a quantum computer.
        Quantum algorithms like HHL achieve quantum speedups with a sophisticated use of these two encodings.
        In this work, we present algorithms that converts these two encodings to one another.
        While quantum digital-to-analog conversions have implicitly been used in existing quantum algorithms, we reformulate it and give a generalized protocol that works probabilistically.
        On the other hand, we propose a deterministic algorithm that performs a quantum analog-to-digital conversion.
        These algorithms can be utilized to realize high-level quantum algorithms such as a nonlinear transformation of amplitudes of a quantum state.
        As an example, we construct a ``quantum amplitude perceptron'', a quantum version of the neural network, and hence has a possible application in the area of quantum machine learning.
    \end{abstract}

    \pacs{Valid PACS appear here}
    \maketitle

    \section{Introduction}
        A wide variety of quantum algorithms that potentially give quantum speedups over classical computers has been proposed.
        The problems that are efficiently solved by existing quantum algorithms can be divided into two types: ones where input data of a problem is relatively small in size but the problem itself is hard classically, and ones where the input of a problem is exponentially large, making it hard to handle on classical computers.
        The formers are solved by algorithms such as Shor's factoring \cite{Shor1997}, or quantum chemistry calculations \cite{Kassal2010}.
        On the other hand, there are algorithms that solve problems categorized as the latter. 
        They achieve quantum speedups only if an oracle that encodes \(N\) classical data in \(O(\poly (\log N))\) time exists \cite{Aaronson2015}.
        A famous example is Harrow-Hassidim-Lloyd (HHL) algorithm \cite{Harrow2009}, which is an algorithm to apply a inverse \(A^{-1}\) of a matrix \(A\) to a \(N=2^n\)-dimensional vector \(\{c_j\}_{j=1}^N\).
        It requires us to construct a quantum state: 
        \begin{equation}\label{analog-encoding}
            \sum_{j=1}^N c_j\ket{j},
        \end{equation}
        where \(\{c_j\}_{j=1}^N\) normalized to satisfy \(\sum_{j=1}^N |c_j|^2 =1\).
        Data encoding in the format of Eq.~(\ref{analog-encoding}) is crucial for all HHL-based algorithms \cite{Schuld2016,Yu2017,Rebentrost2014,Wiebe2012}, and others \cite{Wiebe2015,Lloyd2013}.
        We will call the state of Eq.~(\ref{analog-encoding}) as an ``analog-encoded" state, since data are encoded into analog quantities, that is, complex amplitudes of a quantum state.
        Here we define an analog-encoding unitary transformation \(U_A(\{c_j\})\) by:
        \begin{equation}\label{analog-encoding-unitary}
            U_A(\{c_j\})\ket{0} = \sum_j c_j\ket{j}.
        \end{equation}

        Another approach is to encode \(m\)-bits of binary data into qubit-strings. Let \(N\) and \(\bv{d}_j = \{d_j^{(k)}\}_{k=1}^{m}\) (\(d_j^{(k)}=0,1\), \(j=1,\cdots,N\)) be the number of binary data provided and the data bitstrings. In this approach, data are encoded as follows:
        \begin{equation}\label{digital-encoding}
            \frac{1}{\sqrt{N}}\sum_{j=1}^N \ket{j}\ket{\bv{d}_j} = \frac{1}{\sqrt{N}}\sum_{j=1}^N \ket{j}\ket{d^{(m)}_j\cdots d^{(1)}_j}.
        \end{equation}
        We will call this state as a ``digital-encoded" state.
        For example, quantum algorithms for solving semidefinite programs (SDPs) \cite{Brandao2017a,Brandao2017b} depends on this encoding.
        Similarly to an analog-encoding unitary transformation, we define a digital-encoding unitary transformation by
        \begin{equation}\label{digital-encoding-unitary}
            U_D(\{\bv{d_j}\})\ket{j}\ket{0} = \ket{j}\ket{\bv{d}_j}.
        \end{equation}
        \(U_D\) is often called quantum random access memory (QRAM). 
        Refs. \cite{Giovannetti2008a,Giovannetti2008b} have provided a protocol which employs qutrits to speedup a memory call to \(O(\log^2 N)\) two-body interaction gates.
        Their method is promising compared to a conventional method that requires \(O(N)\) operations.
        It might be worth noting that in the context of QRAM, it is usually assumed that we already have ``memory cells'' which stores data in the form accessible from a quantum computer.
        The \(O(\log^2 N)\) operations do not include the construction of them.
        
        Many quantum algorithms sophisticatedly use these two types of encodings.
        For example, in HHL algorithm \cite{Harrow2009}, an analog-encoded state Eq.~(\ref{analog-encoding}) is put through quantum phase estimation algorithm that digitally encode eigenvalues \(\{\lambda_j\}\) of a matrix \(A\), and then the inverse of them are multiplied to the amplitudes by controlled rotations (Fig.~\ref{fig:sketch} (b));
        in quantum metropolis sampling \cite{Temme2011}, energy eigenvalue \(\{E_j\}\) of a Hamiltonian is first digitally encoded by the phase estimation, and then the encoded energies are transferred to amplitudes in the form of \(e^{-\beta(E_j-E_k)}\), again using controlled rotations. 

        In this paper, we investigate the relation between these two different encoding methods.
        Specifically, we concentrate on conversions between these two encodings;
        can you go from ``digital-encoding" to ``analog-encoding" (quantum digital-to-analog conversion, QDAC), or the other way around (quantum analog-to-digital conversion, QADC)?
        DAC and ADC play important roles in classical information processing, since digitally-stored data are easier to handle than analog data which physical systems generate and are driven by.
        QDAC and QADC can be regarded as quantum analogs of them, and therefore there is a possibility that they stimulate the construction of more sophisticated quantum algorithms.
        
        First, we formulate these problems.
        It is shown that QDAC can be implemented probabilistically, and QADC deterministically.
        A special case of QDAC, in fact, has implicitly been employed in existing algorithms such as HHL and quantum metropolis samplings.
        We unify those techniques and give a generalized procedure.
        QDAC and QADC algorithms provide an insight into what should be done in digital- or analog-encodings.
        Also, as an application, we show that a QADC-QDAC combined method can be utilized to perform almost arbitrary nonlinear transformations of amplitudes of a quantum state.
        This result can be utilized, for example, for a purpose of constructing quantum machine learning algorithms.

        \begin{figure}\label{fig:sketch}
            \includegraphics[width=\linewidth]{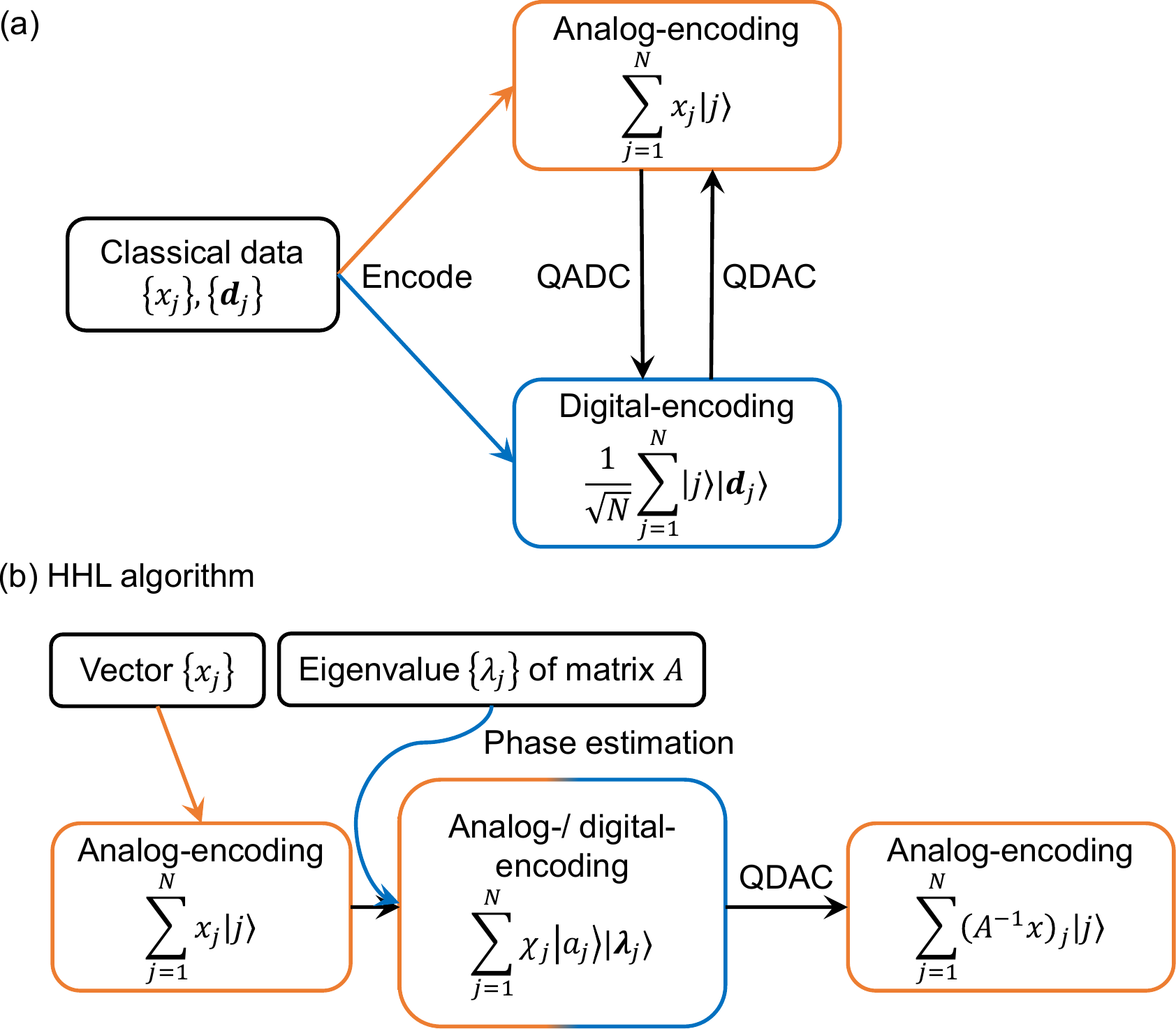}
            \caption{\label{fig:sketch} (a) Schematic sketch of analog-encoding and digital-encoding. QDAC and QADC mediates these two encodings. (b) A brief flowchart of HHL algorithm \cite{Harrow2009}. \(\{\ket{a_j}\}\) denote eigenvectors of a Hermitian matrix \(A\), each corresponding to eigenvalues \(\{\lambda_j\}\). \({\chi_j}\) are complex numbers such that \(\sum_{j=1}^N x_j\ket{j} = \sum_{j=1}^N \chi_j\ket{a_j}\).} 
        \end{figure}

        This paper is organized as follows.
        First in Sec. \ref{sec:preliminary}, we summarize the algorithms we use as subroutines and define QADC and QDAC problem.
        In Sec.\ref{sec:QDAC}, we review QDAC procedures that have implicitly been utilized in existing algorithms.
        We also present the generalized QDAC procedure.
        Then in Sec.\ref{sec:QADC}, we present an algorithm to perform QADC.
        In Sec. \ref{sec:application}, we provide applications of QADC, and show that QADC combined with QDAC provides us a way to perform a nonlinear transformation of amplitudes of a quantum state.

    \section{Preliminary}\label{sec:preliminary}
         Here we summarize some useful results from existing works along with definitions of terms. Throughout this paper \(N=2^n\) denotes the number of data.

        \begin{fact}[analog encoding unitary \cite{Kerenidis2016,Kerenidis2017}]\label{thm:analogencoding}
            For given classical data \(\{x_j\}_{j=1}^N \in \mathbb{R}^N\) such that \(\sum_{j=1}^N x_j^2 = 1\), a binary-tree like classical data structure can be constructed in time \(O(N\log^2 N)\) on a classical computer. With this structure, there exists a quantum algorithm that constructs an analog-encoding unitary \(U_A(\{x_j\}_{j=1}^N)\) with \(O(\log N\poly(\log\log N))\) single- and two-qubit gates.
        \end{fact}
        
        We use the phase estimation algorithm stated below as a key ingredient of our QADC algorithm. 
        
        \begin{fact}[phase estimation \cite{Cleve1997}]\label{thm:phaseestimation}
            Let \(U\) be a unitary operator acting on \(M\)-qubit Hilbert space with eigenstates \(\{\ket{\psi_j}\}_{j=1}^{2^M}\) and corresponding eigenvalues \(\{e^{2\pi i\phi_j}\}_{j=1}^{2^M}\) where \(\phi_j\in [0,1)\). Let \(\epsilon = 2^{-m}\) for some positive integer \(m\). There exists a quantum algorithm, which consists of \(O(1/\epsilon)\) controlled-\(U\) calls and \(O(\log^2(1/\epsilon))\) single- and two-qubit gates, that performs transformation \(\sum_{j=1}^{2^M}a_j\ket{\psi_j}\ket{0}^{\otimes m} \to \ket{\psi_{PE}} = \sum_{j=1}^{2^M}a_j\ket{\psi_j}\ket{\tilde{\bv{\phi}_j}}\) where \(\tilde{\bv{\phi}_j}\) denotes a bitstring \(\tilde{\phi_j}^{(1)}\tilde{\phi_j}^{(2)}\cdots\tilde{\phi_j}^{(m)}\) such that \(|\sum_{k=1}^{m}\tilde{\phi}_j^{(k)} 2^{-k} - \phi_j|\leq \epsilon\) for all \(j\) with state fidelity at least \(1-\poly(\epsilon)\).
        \end{fact}
        We say that \(\ket{\tilde{\psi}}\) has fidelity \(1-\delta\) with \(\ket{\psi}\) when \(|\braket{\tilde{\psi}|\psi}|= 1-\delta\).
        The phase estimation algorithm can also be viewed as a digital-encoding unitary transformation, where the address is replaced by eigenstates of \(U\).

        Next we state a version of the amplitude amplification technique.
        \begin{fact}[amplitude amplification \cite{Brassard2000}]
            Suppose we have a unitary operator \(U\) that acts on \(M\)-qubit Hilbert space as \(U\ket{0}^{\otimes M} = \alpha\ket{\psi}\ket{0} + \beta\ket{G}\ket{1}\) where \(\ket{\psi}, \ket{G}\) are arbitrary \(M-1\)-qubit states. Then probability of getting \(\ket{\psi}\ket{0}\) can be made close to unity by \(O(1/|\alpha|)\) application of \(U\).
        \end{fact}

        We define QDAC as follows.
        
        \begin{definition}[QDAC]
            Let \(\{d_j\}_{j=1}^{N}\) be a set of real numbers in \([0,1)\), each of which is represented by \(d_j = \sum_{k=1}^m d_j^{(k)}2^{-k}\) with binary variables \(d_j^{(k)} \in \{0,1\}\). Let \(\bv{d}_j\) denote the \(m\)-bit string \(d_j^{(1)}\cdots d_j^{(m)}\). An \(m\)-bit QDAC operation transforms digital-encoded state \(\frac{1}{\sqrt{N}}\sum_{j=1}^N \ket{j}\ket{\bv{d}_j}\) to \(C\sum_{j=1}^N d_j\ket{j}\ket{0}^{\otimes m}\) where \(C\) is a normalization constant.
        \end{definition}

        As for QADC, the amplitudes \(\{c_j\}_{j=1}^N\) of a quantum state are, in general, complex numbers. Therefore we define three versions of QADC, each corresponding to the analog-to-digital conversion of the absolute value, the real part, and the imaginary part of \(\{c_j\}\).
        \begin{definition}[abs-QADC]
            Let \(\tilde{\bv{r}_j}\) denote the \(m\)-bit string \(\tilde{r}_j^{(1)}\cdots \tilde{r}_j^{(m)}\) that best approximates \(|c_j|\) by \(\sum_{k=1}^{m}\tilde{r}_j^{(k)} 2^{-k}\). An \(m\)-bit abs-QADC operation transforms analog-encoded state \(\sum_{j=1}^N c_j\ket{j}\ket{0}^{\otimes m}\) to \(\frac{1}{\sqrt{N}}\sum_{j=1}^N \ket{j}\ket{\tilde{\bv{r}}_j}\).
        \end{definition}
        \begin{definition}[real-QADC]
            Let \(\tilde{\bv{x}_j}\) denote the \(m\)-bit string \(\tilde{x}_j^{(1)}\cdots \tilde{x}_j^{(m)}\) that best approximates the real part of \(c_j\) by \(\sum_{k=1}^{m}\tilde{x}_j^{(k)} 2^{-k}\). An \(m\)-bit real-QADC operation transforms analog-encoded state \(\sum_{j=1}^N c_j\ket{j}\ket{0}^{\otimes m}\) to \(\frac{1}{\sqrt{N}}\sum_{j=1}^N \ket{j}\ket{\tilde{\bv{x}}_j}\).
        \end{definition}
        \begin{definition}[imag-QADC]
            Let \(\tilde{\bv{y}_j}\) denote the \(m\)-bit string \(\tilde{y}_j^{(1)}\cdots \tilde{y}_j^{(m)}\) that best approximates the imaginary part of \(c_j\) by \(\sum_{k=1}^{m}\tilde{y}_j^{(k)} 2^{-k}\). An \(m\)-bit imag-QADC operation transforms analog-encoded state \(\sum_{j=1}^N c_j\ket{j}\ket{0}^{\otimes m}\) to \(\frac{1}{\sqrt{N}}\sum_{j=1}^N \ket{j}\ket{\tilde{\bv{y}}_j}\).
        \end{definition}

        We use quantum arithmetics as a subroutine, which is stated as the following theorem.
        \begin{fact}[quantum arithmetics \cite{Ruiz-Perez2017}]
            Let \(\bv{a},\bv{b}\) be \(m\)-bit strings. There exists a quantum algorithm that performs transformation \(\ket{\bv{a}}\ket{\bv{b}} \to \ket{\bv{a}}\ket{\bv{a}+\bv{b}}\) or \(\ket{\bv{a}}\ket{\bv{b}} \to \ket{\bv{a}}\ket{\bv{a}\bv{b}}\) with \(O(\poly(m))\) single- and two-qubit gates.
        \end{fact}
        Note that for accuracy defined as \(\epsilon = 2^{-m}\), quantum arithmetics scales as \(O(\poly(\log (1/\epsilon)))\). Furthermore, we assume the following statement as a fact.
        \begin{fact}\label{thm:arithmetics}
            Some basic functions such as inverse, trigonometric functions, square root, and inverse trigonometric functions can be calculated to accuracy \(\epsilon\), that is, we can perform a transformation \(\ket{\bv{a}}\ket{\bv{0}}\to\ket{\bv{a}}\ket{\tilde{\bv{f}}(\bv{a})}\) such that \(|\tilde{f}(a)-f(a)|\leq \epsilon\) where \(f(a)\) is the objective function, using \(O(\poly(\log(1/\epsilon)))\) quantum arithmetics.
        \end{fact}

        A similar title is found on Ref.~\cite{Schmuser2005}.
        However, their purpose was to map a continuous-space wave function \(\ket{\psi}=\int dx \psi(x)\ket{x}\) to a discrete-space wave function \(\ket{\psi_d} = \sum_j \psi(j/N)\ket{j}\). In the context of this paper, it can be viewed as analog-to-analog conversion.
    
    \section{QDAC}\label{sec:QDAC}
        It is actually straightforward to create an analog-encoded state from a digital-encoded state just by adding an ancilla qubit and performing a controlled rotation with the data register.
        In fact, this QDAC procedure has implicitly used in existing works.
        For example, HHL \cite{Harrow2009} has utilized the above protocol to multiply the inverse of eigenvalues of a Hermitian matrix \(A\) to an analog-encoded state vector.
        We state this QDAC operation formally as a theorem below.
        \begin{theorem}[QDAC with ancilla]
            There exists a quantum algorithm that performs \(m\)-bit QDAC using \(O(\poly(\log(1/\epsilon)))\) single- and two-qubit gates and one \(U_D^\dagger\), where \(\epsilon=2^{-m}\), with probability \(\sum_{j=1}^{N}d_j^2/N\).
        \end{theorem}
        {\it Proof -} The procedure of the algorithm is as follows. Assume that we are provided with a digital-encoded state Eq.~(\ref{digital-encoding}).
        \begin{enumerate}
            \item Compute \(\varphi_j = \frac{2}{\pi}\cos^{-1}d_j\) by quantum arithmetics.
            \begin{equation}\label{eq:arccoscalc}
                \frac{1}{\sqrt{N}}\sum_{j=1}^N \ket{j}\ket{\bv{d}_j}\ket{0}^{\otimes m}
                \to \frac{1}{\sqrt{N}}\sum_{j=1}^N \ket{j}\ket{\bv{d}_j}\ket{\bv{\varphi}_j},
            \end{equation}
            where \(\bv{\varphi}_j\) denotes \(m\)-bit strings \(\varphi_j^{(1)}\cdots \varphi_j^{(m)}\) such that \(\varphi_j = \sum_{k=1}^{m} \varphi_j^{(k)}2^{-k}\).
            \item Adding an ancilla qubit \(\ket{0}_a\), perform a controlled rotation \(R_y(\pi\varphi_j) = e^{i\pi\varphi_j Y/2}\), on the ancilla.
            \begin{align}\label{eq:controlled_rotation}
                &\frac{1}{\sqrt{N}}\sum_{j=1}^N \ket{j}\ket{\bv{d}_j}\ket{\bv{\varphi}_j}\ket{0}_a\nonumber \\
                &\to \frac{1}{\sqrt{N}}\sum_{j=1}^N \ket{j}\ket{\bv{d}_j}\ket{\bv{\varphi}_j}\left(d_j\ket{0}_a + \sqrt{1-{d_j^2}}\ket{1}_a\right).
            \end{align}
            \item Measure the ancilla in computational basis. With probability \(\sum_{j=1}^N d_j^2/N\), we obtain
            \begin{equation}
                C\sum_{j=1}^N d_j\ket{j}\ket{\bv{d}_j}\ket{\bv{\varphi}_j}\ket{0}_a,
            \end{equation}
            where \(C = \sqrt{1/(\sum_{j=1}^N d_j^2)}\).
            \item Uncompute \(\varphi_j\) (step 1) and apply \(U_D^\dagger\). We now have an analog-encoded state
            \begin{equation}
                C\sum_{j=1}^N d_j\ket{j}.
            \end{equation}
        \end{enumerate}
        Now we analyze the complexity.
        On step 1, \(\cos^{-1} d_j\) can be calculated with quantum arithmetics using \(O(\poly(\log(1/\epsilon)))\) gates by Fact~\ref{thm:arithmetics}.
        Step 2 uses \(m = \log (1/\epsilon)\) controlled rotations.
        Therefore, overall complexity for steps 1, 2 and 4 is \(O(\poly(\log(1/\epsilon)))\) and one \(U_D^\dagger\).
        The success probability of step 3 is \(\sum_{j=1}^N d_j^2/N\). \(\hfill\square\)
        
        Success probability of above procedure can be rewritten in terms of the mean \(\mu\) and the variance \(v\) of the data, since \(\frac{1}{N}\sum_{j=1}^{N}d_j^2 = v+\mu^2\).
        Note that, when \(v+\mu^2 \ll 1\) and the success probability is relatively small, the amplitude amplification technique can be utilized to shorten the expected running time quadratically from \(O(1/(v+\mu^2))\) to \(O\left(1/\sqrt{(v+\mu^2)}\right)\).
        
        When one modifies Step 1 to give \(\varphi_j = \cos^{-1}(f(d_j))\), we readily obtain the following.
        \begin{corollary}[generalized QDAC]\label{col:QDAC}
            There exists a quantum algorithm that performs the transformation \(\frac{1}{\sqrt{N}}\sum_{j=1}^N \ket{j}\ket{\bv{d}_j} \to C'\sum_{j=1}^N \tilde{f}(d_j)\ket{j}\ket{0}\) such that \(|\tilde{f}(d_j)-f(d_j)|\leq\epsilon\), where \(f: [0,1)\to[0,1)\) is a function satisfying Fact~5, using \(O(\poly(\log(1/\epsilon)))\) single- and two-qubit gates, with probability \(\sum_{j=1}^{N}\tilde{f}(d_j)^2/N = (NC')^{-1}\).
        \end{corollary}
        If we choose \(f(x) = \tanh(x)\), Collorary~\ref{col:QDAC} provides an alternative way to implement a sigmoid function other than the one proposed in Ref.~\cite{Cao2017}.

    \section{QADC}\label{sec:QADC}
\begin{figure}
            \includegraphics[width=0.7\linewidth]{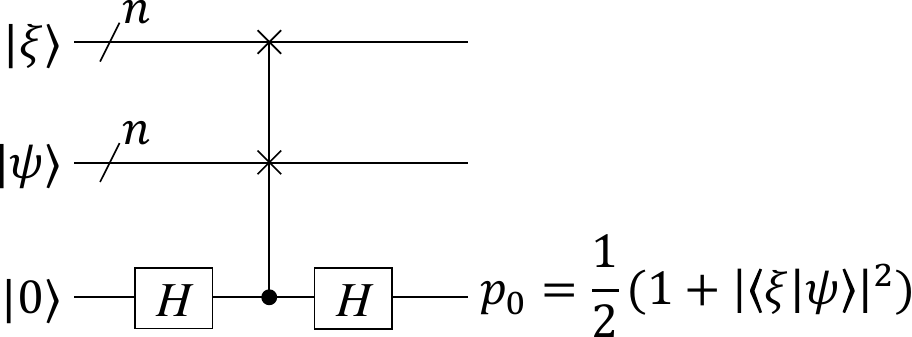}
            \caption{\label{fig:swap_test} Quantum circuit of the swap test \cite{Buhrman2001}.}
        \end{figure}

\begin{figure}
            \includegraphics[width=\linewidth]{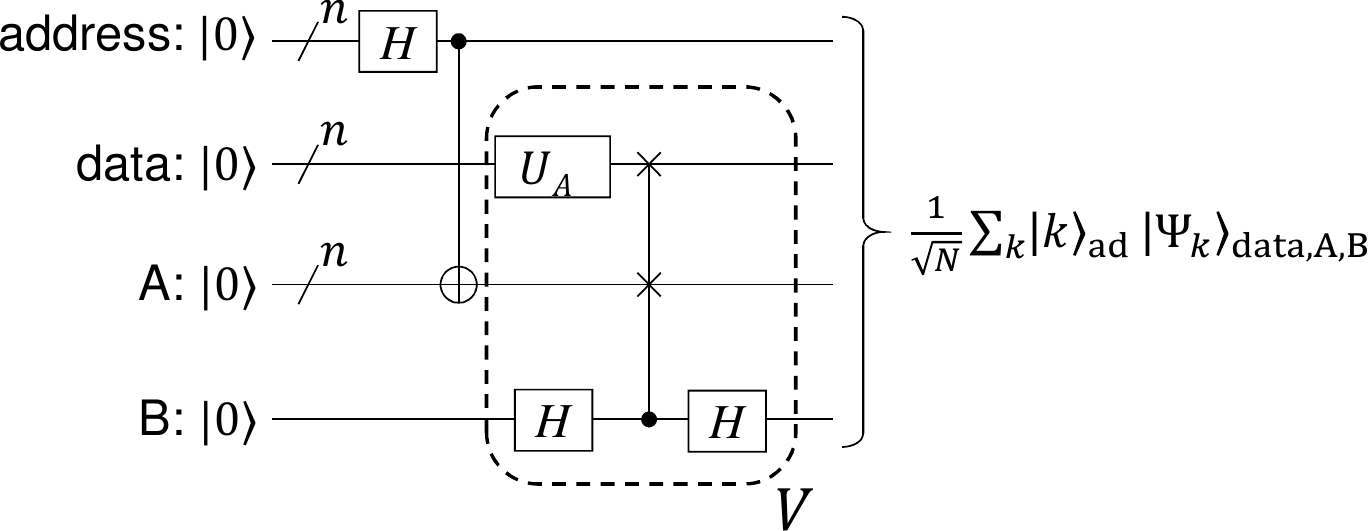}
            \caption{\label{fig:toswaptest} Quantum circuit through steps 1 to 4 of abs-QADC in the main text.}
        \end{figure}

\begin{figure}
            \includegraphics[width=0.8\linewidth]{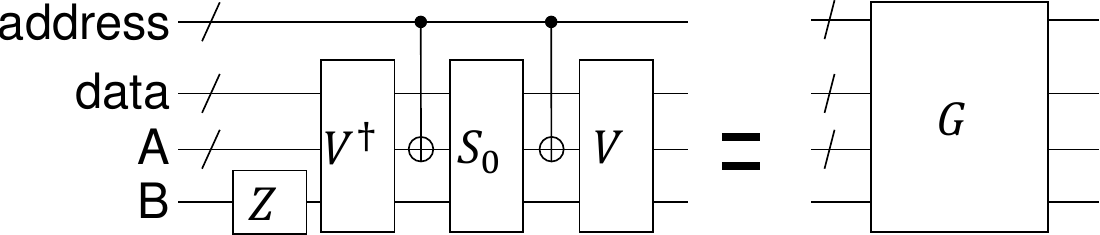}
            \caption{\label{fig:groveriteration} Definition of gate \(G\) in abs-QADC.}
        \end{figure}

\begin{figure*}
            \includegraphics[width=0.8\linewidth]{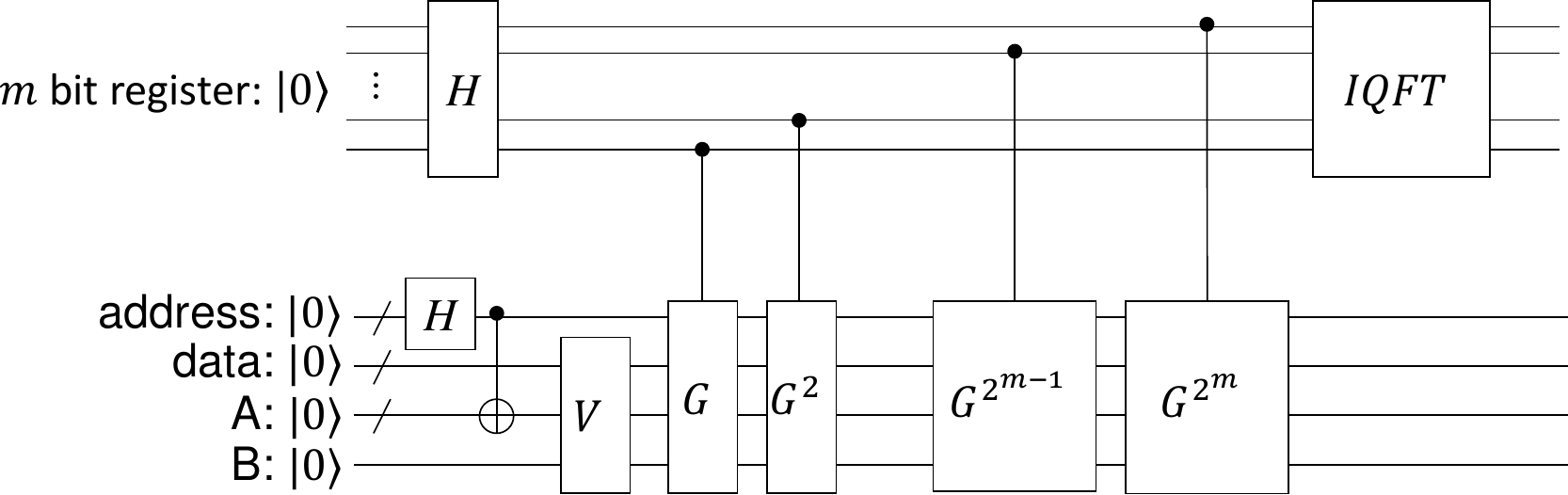}
            \caption{\label{fig:AE} Step 6 of the abs-QADC algorithm. The phase estimationa is performed to encode the analog-encoded value \(x_j\) into qubit bitstrings. IQFT is inverse quantum Fourier transformation \cite{Nielsen2010}.}
        \end{figure*}

        First, we propose an abs-QADC algorithm.
        Note that the abs-QADC may easily be constructed with the real-QADC and the imag-QADC presented as Theorem \ref{thm:real-QADC} and \ref{thm:imag-QADC}.
        However, we expect that the abs-QADC algorithm that we present here would bring you some intuitions in the construction of the algorithm.
        We use the swap test \cite{Buhrman2001}, which is a special case of the Hadamard test, to extract the absolute value of amplitudes.
        The usual swap test, as described in Fig.~\ref{fig:swap_test}, measures an absolute value of an inner product of arbitrary two states \(\ket{\psi}\) and \(\ket{\xi}\) as \(p_0\), the probability of getting \(\ket{0}\) from an ancilla qubit.
        If we input \(\ket{k}\), which is a computational basis state, and an analog-encoded state to the swap test, we can extract a data \(x_k\).
        The amplitude estimation \cite{Brassard2000} of \(p_0\) can be utilized to encode the data digitally.
        An important trick used in the algorithm presented below is that this process can be parallelized.
        \begin{theorem}[abs-QADC]\label{thm:abs-QADC}
            There exists an \(m\)-bit abs-QADC algorithm that runs using \(O(1/\epsilon)\) controlled-\(U_A\) gates and \(O((\log^2 N)/\epsilon)\) single- and two-qubit gates with output state fidelity \(1-O(\poly(\epsilon))\), where \(\epsilon=2^{-m}\).
        \end{theorem}
        {\it Proof - }First we provide the algorithm. (See Fig.~\ref{fig:toswaptest} for steps 1-4.)
        \begin{enumerate}
            \item Prepare address qubits. \(\frac{1}{\sqrt{N}}\sum_{k=1}^N \ket{k}_{\text{ad}}\)
            \item Perform controlled-NOT from the address qubits to initialized ancilla qubits, which will be referred as qubits A, to get \(\frac{1}{\sqrt{N}}\sum_{k=1}^N \ket{k}_{\text{ad}}\ket{k}_A\).
            \item Prepare the analog-encoded state in data qubits, \(\sum_j c_j\ket{j}_{\text{data}}\).
            \item Using another ancilla qubit (We will call it qubit B), perform a swap test \cite{Buhrman2001} without measurement between data-qubit and qubits A (Fig.~\ref{fig:toswaptest}). We have:
            \begin{align}
                &\sum_k \frac{\ket{k}_{\text{ad}}}{2\sqrt{N}} \left[\left(\sum_j c_j\ket{j}_{\text{data}} \ket{k}_A + \ket{k}_{\text{data}}\sum_j c_j\ket{j}_{A}\right)\ket{0}_B + \right.\nonumber\\
                &\quad \left.\left(\sum_j c_j\ket{j}_{\text{data}} \ket{k}_A - \ket{k}_{\text{data}}\sum_j c_j\ket{j}_{A}\right)\ket{1}_B\right]\\
                &\equiv \frac{1}{\sqrt{N}}\sum_k \ket{k}_{\text{ad}}\ket{\Psi_k}_{\text{data},A,B}.
            \end{align}
            Fig.~\ref{fig:toswaptest} shows the quantum circuit from step 1 to step 4.
            We define \(V\) to be the combined unitary transformation of step 3 and 4.
            This step extracts an absolute value \(r_k\) of amplitude \(c_k\) each corresponding to an address \(\ket{k}_{\text{ad}}\).
            The similar idea is also used in Ref.~\cite{Zhou2017}.
            \item Construct a gate 
            \begin{equation}\label{eq:G}
                G = V (\text{CNOT})_{\text{ad}\to A} S_0 (\text{CNOT})_{\text{ad}\to A} V^\dagger Z_B,
            \end{equation}
            where \(S_0\) is a conditional phase shift gate; \(S_0 = I - 2\left(\ket{0}\bra{0}\right)_{\text{data,A,B}}\) and \(Z_B\) is a Pauli \(Z\) gate only acting on the qubit B (Fig.~\ref{fig:groveriteration}).
            The act of \(G\) can be written as
            \begin{align}\label{eq:G_k}
                &G \frac{1}{\sqrt{N}}\sum_k \ket{k}_{\text{ad}}\ket{\Psi_k}_{\text{data},A,B} = \nonumber\\ &\quad\frac{1}{\sqrt{N}}\sum_k \ket{k}_{\text{ad}}\left(G_k \ket{\Psi_k}_{\text{data},A,B}\right),
            \end{align}
            where
            \begin{equation}
                G_k = V S_k V^\dagger Z_B,
            \end{equation}
            and 
            \begin{equation}
                S_k = I - 2\left(\ket{0}\bra{0}\right)_{\text{data,B}}\otimes (\ket{k}\bra{k})_A.
            \end{equation}
            Each \(\ket{\Psi_k}_{\text{data},A,B}\) is decomposed into two of eigenstates \(\ket{\Psi_{k+}}_{\text{data},A,B}\) and \(\ket{\Psi_{k-}}_{\text{data},A,B}\) of \(G_k\), each respectively corresponding to eigenvalue \(\lambda_{k\pm} = e^{\pm i2\pi\theta_k}\) where \(\sin(\pi\theta_k) = \sqrt{\frac{1}{2}(1+r_k^2)}\) and \(\theta_k \in [1/4,1/2)\).
            The decomposition is: \(\ket{\Psi_k}_{\text{data},A,B} = \frac{-i}{\sqrt{2}}(e^{i\pi\theta_k}\ket{\Psi_{k+}}_{\text{data},A,B}-e^{-i\pi\theta_k}\ket{\Psi_{k-}}_{\text{data},A,B})\). See Appendix for detailed description.

            \item Introducing the register qubits, run the phase estimation of \(G\) as depicted in Fig.\ref{fig:AE}. Then we have:
            \begin{align}
                &\frac{1}{\sqrt{2N}}\sum_k \ket{k}_{\text{ad}}\left(\ket{\bv{\theta}_k}_{\text{reg}'}\ket{\Psi_{k+}}_{\text{data},A,B}\right.\nonumber\\
                &\qquad\left.+\ket{\bv{1-\theta}_k}_{\text{reg}'}\ket{\Psi_{k-}}_{\text{data},A,B}\right) \nonumber \\
                &\equiv \frac{1}{\sqrt{N}}\sum_k \ket{k}_{\text{ad}}\ket{\Psi_{k,AE}}_{\text{reg}',\text{data},A,B}
            \end{align}
            where \(\ket{\bv{\theta}_k}_{\text{reg}'}\) and \(\ket{\bv{1-\theta}_k}_{\text{reg}'}\) are \(m\)-bit strings that store \(\theta_k\) and \(1-\theta_k\) as binary data, and 
            \begin{align}
                &\ket{\Psi_{k,AE}}_{\text{reg}',\text{data},A,B} = \nonumber \\
                 &\quad \frac{1}{\sqrt{2}}\left(\ket{\bv{\theta}_k}_{\text{reg}'}\ket{\Psi_{k+}}_{\text{data},A,B}+\ket{\bv{1-\theta}_k}_{\text{reg}'}\ket{\Psi_{k-}}_{\text{data},A,B}\right)
            \end{align}

            \item On another register, using digital quantum arithmetics, calculate \(r_k = \sqrt{2\sin^2 \pi\theta_k-1}\). Note that \(\sin\pi\theta_k = \sin\pi(1-\theta_k)\), and \(r_k\) is uniquely recovered since \(r_k\in[0,1]\). Then finally we get:
            \begin{align}
                \frac{1}{\sqrt{N}}\sum_{k=0}^N \ket{k}_{\text{ad}}\ket{\tilde{\bv{r}}_k}_{\text{reg}}\ket{\Psi_{k,AE}}_{\text{reg}',\text{data},A,B}.
            \end{align}
            \item Uncompute the data, A, B, reg' qubits. We obtain:
            \begin{equation}
                \frac{1}{\sqrt{N}}\sum_{k=0}^N \ket{k}_{\text{ad}}\ket{\tilde{\bv{r}}_k}_{\text{reg}}\ket{0}_{\text{reg}',\text{data},A,B},
            \end{equation}
            which is a digital-encoded state.
        \end{enumerate}
        
        Here we analyze the complexity of the above algorithm.
        For steps 1 to 4, we used \(O(\log N)\) single- and two-qubit gate.
        On step 5, the phase estimation, we need to use \(O(1/\epsilon)\) of controlled-\(U_A\) and \(O(\log^2 N/\epsilon)\) of single- and two-qubit gate.
        Step 6, quantum arithmetics takes \(O(\poly(\log(1/\epsilon)))\) by Fact~5.
        Therefore, overall complexity is \(O(1/\epsilon)\) of controlled-\(U_A\) and \(O(\log^2 N/\epsilon)\) of single- and two-qubit gate.
        The fidelity of the output state is \(1-O(\poly(\epsilon))\) by Fact~\ref{thm:phaseestimation}.~\(\hfill\square\)
        
\begin{figure}
            \includegraphics[width=\linewidth]{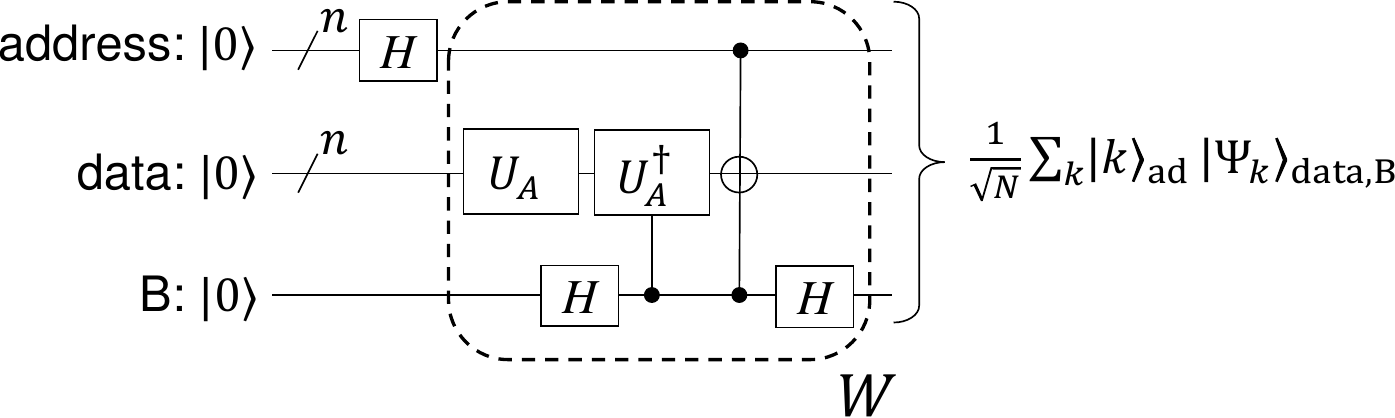}
            \caption{\label{fig:hadamardtest_real} A quantum circuit through steps 1 to 3 of real-QADC in the main text.}
        \end{figure}

        Next we show the real-QADC.
        \begin{theorem}[real-QADC]\label{thm:real-QADC}
            There exists an \(m\)-bit real-QADC algorithm that runs using \(O(1/\epsilon)\) controlled-\(U_A\) gates and \(O((\log^2 N)/\epsilon)\) single- and two-qubit gates with output state fidelity \(1-O(\poly(\epsilon))\), where \(\epsilon=2^{-m}\).
        \end{theorem}
        {\it Proof - }We provide the algorithm. (See Fig.~\ref{fig:toswaptest} for steps 1-3.) The algorithm presented here is slightly modified one from the previous algorithm for QADC.
        \begin{enumerate}
            \item Prepare address qubits. \(\frac{1}{\sqrt{N}}\sum_{k=1}^N \ket{k}_{\text{ad}}\)
            \item Prepare the analog-encoded state in data qubits, \(\sum_j x_j\ket{j}_{\text{data}}\)
            \item Using another ancilla qubit (We will call it qubit B), perform an Hadamard test as described in Fig.~\ref{fig:hadamardtest_real}. We have:
            \begin{align}
                &\sum_k \frac{\ket{k}_{\text{ad}}}{2\sqrt{N}} \left[\left(\sum_j x_j\ket{j}_{\text{data}} + \ket{k}\right)\ket{0}_B + \right.\nonumber\\
                &\quad \left.\left(\sum_j x_j\ket{j}_{\text{data}} - \ket{k}\right)\ket{1}_B\right]\\
                &\equiv \frac{1}{\sqrt{N}}\sum_k \ket{k}_{\text{ad}}\ket{\Psi_k}_{\text{data},B}.
            \end{align}
            This step extracts the real part \(x_k\) of a complex amplitude \(c_k\) each corresponding to an address \(\ket{k}_{\text{ad}}\).
            Fig.~\ref{fig:hadamardtest_real} shows the quantum circuit from step 1 to 3.
            We define \(W\) to be the combined unitary transformation of step 2 and 3.
            \item Construct a gate 
            \begin{equation}\label{eq:G_real}
                G' = W S_0' W^\dagger Z_B,
            \end{equation}
            where \(S_0'\) is a conditional phase shift gate; \(S_0' = I - 2\left(\ket{0}\bra{0}\right)_{\text{data,B}}\) and \(Z_B\) is a Pauli \(Z\) gate only acting on the qubit B (Fig.~\ref{fig:groveriteration}).
            The act of \(G'\) can be written as
            \begin{align}\label{eq:G_k}
                &G' \frac{1}{\sqrt{N}}\sum_k \ket{k}_{\text{ad}}\ket{\Psi_k}_{\text{data},B} = \nonumber\\ &\quad\frac{1}{\sqrt{N}}\sum_k \ket{k}_{\text{ad}}\left(G_k' \ket{\Psi_k}_{\text{data},B}\right),
            \end{align}
            where
            \begin{equation}
                G_k' = (1-2\ket{\Psi_k}_{\text{data},B}\bra{\Psi_k}_{\text{data},B}) Z_B.
            \end{equation}
            Each \(\ket{\Psi_k}_{\text{data},B}\) is decomposed into two of eigenstates \(\ket{\Psi_{k+}}_{\text{data},B}\) and \(\ket{\Psi_{k-}}_{\text{data},B}\) of \(G_k\), each respectively corresponding to eigenvalue \(\lambda_{k\pm} = e^{\pm i2\pi\theta_k}\) where \(\sin(\pi\theta_k) = \sqrt{\frac{1}{2}(1+x_k)}\) and \(\theta_k \in [1/4,1/2)\).
            The decomposition is: \(\ket{\Psi_k}_{\text{data},B} = \frac{-i}{\sqrt{2}}(e^{i\pi\theta_k}\ket{\Psi_{k+}}_{\text{data},B}-e^{-i\pi\theta_k}\ket{\Psi_{k-}}_{\text{data},B})\).
            The detail of this transformation is similar to the one described in Appendix and thus omitted.

            \item Introducing the register qubits, run the phase estimation of \(G'\). Then we have:
            \begin{align}
                &\frac{1}{\sqrt{2N}}\sum_k \ket{k}_{\text{ad}}\left(\ket{\bv{\theta}_k}_{\text{reg}'}\ket{\Psi_{k+}}_{\text{data},B}\right.\nonumber\\
                &\qquad\left.+\ket{\bv{1-\theta}_k}_{\text{reg}'}\ket{\Psi_{k-}}_{\text{data},B}\right) \nonumber \\
                &\equiv \frac{1}{\sqrt{N}}\sum_k \ket{k}_{\text{ad}}\ket{\Psi_{k,AE}}_{\text{reg}',\text{data},B}
            \end{align}
            where \(\ket{\bv{\theta}_k}_{\text{reg}'}\) and \(\ket{\bv{1-\theta}_k}_{\text{reg}'}\) are \(m\)-bit strings that store \(\theta_k\) and \(1-\theta_k\) as binary data, and 
            \begin{align}
                &\ket{\Psi_{k,AE}}_{\text{reg}',\text{data},B} = \nonumber \\
                 &\quad \frac{1}{\sqrt{2}}\left(\ket{\bv{\theta}_k}_{\text{reg}'}\ket{\Psi_{k+}}_{\text{data},B}+\ket{\bv{1-\theta}_k}_{\text{reg}'}\ket{\Psi_{k-}}_{\text{data},B}\right)
            \end{align}

            \item On another register, using digital quantum arithmetics, calculate \(x_k = 2\sin^2 \pi\theta_k-1\). Note that \(\sin\pi\theta_k = \sin\pi(1-\theta_k)\). Then finally we get:
            \begin{align}
                \frac{1}{\sqrt{N}}\sum_{k=0}^N \ket{k}_{\text{ad}}\ket{\tilde{\bv{x}}_k}_{\text{reg}}\ket{\Psi_{k,AE}}_{\text{reg}',\text{data},B}.
            \end{align}
            \item Uncompute the data, A, B, reg' qubits. We obtain:
            \begin{equation}
                \frac{1}{\sqrt{N}}\sum_{k=0}^N \ket{k}_{\text{ad}}\ket{\tilde{\bv{x}}_k}_{\text{reg}}\ket{0}_{\text{reg}',\text{data},B},
            \end{equation}
            which is a digital-encoded state.
        \end{enumerate}
        The runtime of this algorithm is as same as the abs-QADC. ~\(\hfill\square\)

\begin{figure}
            \includegraphics[width=\linewidth]{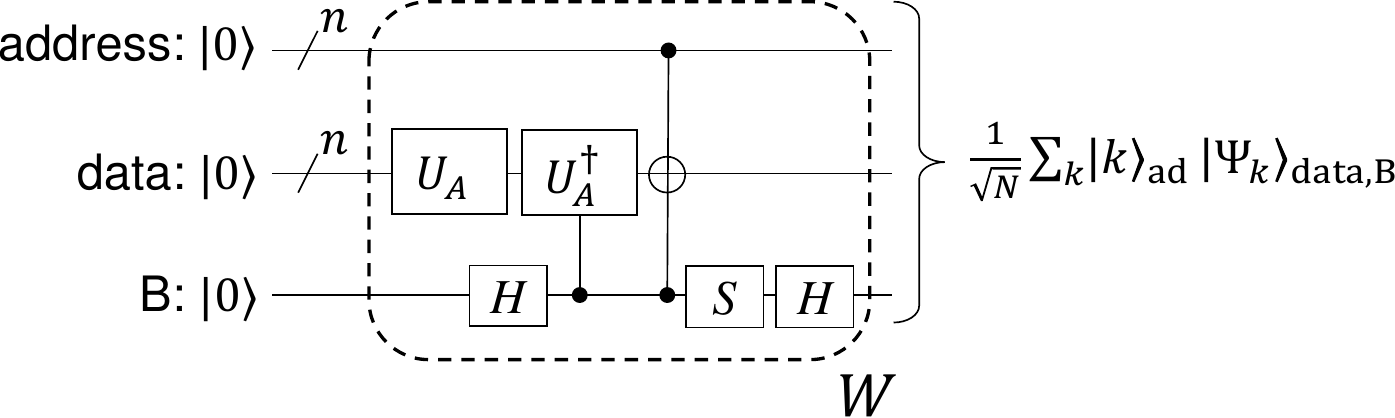}
            \caption{\label{fig:hadamardtest_imag} A quantum circuit element for imag-QADC. The imag-QADC can be performed by replacing \(W\) in the real-QADC by this circuit.}
        \end{figure}

        The imag-QADC can be constructed in the same manner as the real-QADC.
        In this case, we replace the gate \(W\) to the one described in Fig.~\ref{fig:hadamardtest_imag}.
        Therefore we have the following.
        \begin{theorem}[imag-QADC]\label{thm:imag-QADC}
            There exists an \(m\)-bit imag-QADC algorithm that runs using \(O(1/\epsilon)\) controlled-\(U_A\) gates and \(O((\log^2 N)/\epsilon)\) single- and two-qubit gates with output state fidelity \(1-O(\poly(\epsilon))\), where \(\epsilon=2^{-m}\).
        \end{theorem}
        
    \section{applications}\label{sec:application}

    \subsection{Classical data loading}
        As stated in the introduction, there are some algorithms, such as quantum SDP solvers \cite{Brandao2017a,Brandao2017b}, that require an oracle which encodes \(N\) classical data into a quantum state Eq.~(\ref{digital-encoding}) in time \(O(\poly(\log N))\).
        
        QADC can be utilized for this purpose.
        Assume that a binary-tree structure required in Fact~\ref{thm:analogencoding} is already constructed on the classical side.
        Combined with the analog-encoding unitary \(U_A\) of Fact~\ref{thm:analogencoding}, the \(m\)-bit digital-encoded state \(\frac{1}{\sqrt{N}}\sum_{j=1}^N\ket{j}\ket{\bv{d}_j}\) can be prepared with \(O(\log^2 N\poly(\log\log N)/\epsilon)\) quantum gates.
        
        Recently another method for digital encoding has been proposed \cite{Cortese2018}.
        They proposed a protocol that directly encodes binary classical data into qubits.
        
    \subsection{Nonlinear transformation of amplitude}
        The transformation performed on the probability amplitudes \(c_j\) of a quantum state \(\sum_{j=1}^N c_j \ket{j}\), without measurements, is always linear owing to the unitarity of the quantum dynamics.
        Even with (projective) measurement, the transformation is restricted to the form of \(c_j/C\), where C is some constant that is determined by the normalization condition.
        Therefore, a transformation of the form \(f(c_j)\) with an arbitrary function \(f\) cannot be done, without encoding them in digital form using QADC.
        We state the fact that this form of nonlinear transformation of amplitude can be performed using QADC and QDAC as the following theorem.
        \begin{theorem}\label{thm:nonlinear}
            Let data \(\{c_j\}_{j=1}^N\), analog-encoding unitary \(U_A(\{c_j\}_{j=1}^N)\) and analog-encoded state \(\sum_{j=1}^N c_j\ket{j}\) be given. For any function \(f:\mathbb{C} \to [-1,1]\) that satisfies Fact~5, there exists a probabilistic quantum algorithm that performs transformation \(\sum_{j=1}^N c_j\ket{j} \to C'\sum_{j=1}^N \tilde{f}(c_j)\ket{j}\), such that \(|\tilde{f}(c_j)-f(c_j)|\leq\epsilon\), with \(O(1/\epsilon)\) controlled-\(U_A\) gates and \(O((\log^2 N)/\epsilon)\) single- and two-qubit gates. Probability of success is \(\sum_{j=1}^N \tilde{f}(c_j)^2/N\) and the output state fidelity is \(1-O(\poly(\epsilon))\)
        \end{theorem}
        {\it Proof -} By running the QADC algorithm from Theorem~\ref{thm:real-QADC} and Theorem~\ref{thm:imag-QADC} in parallel, we get a state \(\frac{1}{\sqrt{N}}\sum_{j=1}^N \ket{j}\ket{\tilde{\bv{x}}_j}\ket{\tilde{\bv{y}}_j}\) using \(O(1/\epsilon)\) controlled-\(U_A\) gates and \(O((\log^2 N)/\epsilon)\) single- and two-qubit gates with state fidelity \(1-O(\poly(\epsilon))\). Then performing the modified QDAC from Corollary \ref{col:QDAC}, we get \(C'\sum_{j=1}^N \tilde{f}(x_j, y_j)\ket{j}\ket{\bv{x}_j}\ket{\bv{y}_j}\) with \(O(\poly(\log(1/\epsilon)))\) single- and two-qubit gates and with probability \(\sum_{j=1}^N \tilde{f}(c_j)^2/N\). Finally, application of inverse QADC leaves us the desired state. \(\hfill\square\)

        The most useful application of this would be to deal with a ``quantum big-data".
        For example, using \(U_A\) on two registers, we can prepare
        \begin{equation}\label{eq:tensorprod}
            U_A\ket{0}\otimes U_A\ket{0} = \sum_{i,j} c_ic_j\ket{i}\ket{j}.
        \end{equation}
        \(U_A\) is not restricted to the loading of classical data \(\{c_j\}\), but can also be a time evolution operator \(e^{-iHt}\), where \(H\) is some Hamiltonian which can efficiently be simulated on a quantum computer.
        Notice that the tensor product structure of quantum mechanics has calculated the product \(c_i,c_j\) and the combined system has \(N^2\) of amplitudes.
        This nonlinearity is employed in quantum circuit learning \cite{Mitarai2018} for machine learning application.
        The direct digital encoding of these \(N^2\) dimensional data on a quantum state might be impractical, since the calculation of the product \(\{c_ic_j\}_{i,j=1}^N\) takes \(O(N^2)\) time and space classically.
        In comparison, the construction of the quantum state Eq.~(\ref{eq:tensorprod}) requires only \(O(N\log N)\) classical and \(O(\log N\poly(\log\log N))\) quantum operations.
        Tensor product structure is useful to introduce nonlinearity to the data in the form of Eq.~(\ref{eq:tensorprod}), however, it cannot be used to introduce general and more complex transformation stated in above theorem.
        
        Applying QADC on this combined state yields a state,
        \begin{equation}
            \sum_{i,j} \ket{i}\ket{j}\ket{c_i c_j}.
        \end{equation}
        Note that this transformation can also be performed in time \(O(poly(\log N))\).
        Then applying the nonlinear QDAC procedure and the inverse of QADC, one gets,
        \begin{equation}
            \sum_{i,j} f(c_ic_j)\ket{i}\ket{j}\ket{0},
        \end{equation}
        which has nonlinearly transformed probability amplitudes.
        A further extension is discussed in the next subsection.

    \subsection{Quantum amplitude perceptron}
        When one chooses \(f(x)=\tanh (x)\) or the ReLU function which are frequently used in neural networks, it can readily mimic the perceptron, which is a building block of neural networks.
        We denote such an activation function by \(\sigma(x)\)
        In neural network, a transformation of the form \(\sigma(\bv{w}\cdot\bv{x})\), where \(\bv{w}\) and \(\bv{x}\) are a weight vector and a data vector, is utilized to learn some task.
        The training of the network is done by tuning the weight \(\bv{w}\) to give some specific output.
        To mimic this, first we apply a parametrized unitary transformation \(U(\bv{\theta})\), which corresponds to the weight \(\bv{w}\), on an analog-encoded state \(\sum_{i=1}^N c_i \ket{i}\), yielding a state \(\sum_{i=1}^N \sum_{j=1}^N u_{jk}(\bv{\theta})c_j\ket{i}\).
        Note that the amplitude \(x_i\) can be a resultant amplitude after the tensor product multiplication described in the previous section.
        Then use the procedure of Theorem~\ref{thm:nonlinear} with the activation function \(\sigma\), which produces,
        \begin{equation}\label{eq:mimic_perceptron}
            C'\sum_{i} \sigma\left[\sum_ju_{ij}(\bv{\theta})x_j\right]\ket{i}.
        \end{equation}
        The full tomography of this state would require us an exponential time, but if one is interested in the constant number of perceptron outputs, performing a swap test \cite{Buhrman2001} with chosen basis \(\{\ket{k}\}\) is enough to extract them.
        The state Eq.~(\ref{eq:mimic_perceptron}) can be further transformed by some unitary gates to determine the readout weight of the output.
        
        For machine learning, it is necessary to optimize the parameter \(\bv{\theta}\).
        Although the gradient of the output with respect to \(\bv{\theta}\) cannot be extracted due to the complex form of Eq.~(\ref{eq:mimic_perceptron}), gradient-free methods can be utilized, just as same as mentioned in Ref.~\cite{Cao2017}.
        
        Finally, we note here, if one can determine the value of \(\sum_i \left\{\sigma\left[\left(\sum_ju_{ij}(\bv{\theta})x_j\right)^2\right]\right\}^2\) somehow, amplitude amplification technique can be employed to make QDAC procedure deterministic.
        This would enable us to implement a multilayer neural network on a quantum computer deterministically.

    \section{Conclusion}
        We have formulated QDAC and QADC, and described algorithms to implement them on quantum computers.
        Although QDAC protocols have implicitly been utilized in existing algorithms, we have generalized it to facilitate complex nonlinear functions. 
        We have also presented an algorithm that performs QADC.
        It was shown that a combination of QADC and QDAC enables us almost arbitrary nonlinear transformations of amplitudes of a quantum state.
        The possible application of this nonlinear transformation is to make a quantum amplitude perceptron, which can be employed to construct more sophisticated quantum machine learning algorithms.
    
    \begin{acknowledgments}
        KM would like to thank Makoto Negoro for fruitful discussions. KF is supported by KAKENHI No. 16H02211, JST PRESTO JPMJPR1668, JST ERATO JPMJER1601, and JST CREST JPMJCR1673.
        KM and MK are supported by CREST by JST grant number JPMJCR1672.
    \end{acknowledgments}

    \appendix*
    \section{Eigenvalues and eigenstates of \(G_k\)}
    Here we calculate the eigenvalues and eigenstates of \(G_k\) defined in Eq. (\ref{eq:G_k}).
    We consider \(G_k\) acting on the state
    \begin{align}
        & \ket{\Psi_k}_{\text{data}, A, B} = V \ket{0}_{\text{data}}\ket{k}_A\ket{0}_B = \nonumber \\
        &\frac{1}{2} \left[\left(\sum_j c_j\ket{j}_{\text{data}} \ket{k}_A + \ket{k}_{\text{data}}\sum_j c_j\ket{j}_{A}\right)\ket{0}_B + \right.\nonumber\\
        &\quad \left.\left(\sum_j c_j\ket{j}_{\text{data}} \ket{k}_A - \ket{k}_{\text{data}}\sum_j c_j\ket{j}_{A}\right)\ket{1}_B\right].
    \end{align}
    
    First, we define two normalized states
    \begin{align}
        \ket{\Psi_{k0}} &= \frac{1}{2\alpha_k} \left(\sum_j c_j\ket{j}_{\text{data}} \ket{k}_A + \ket{k}_{\text{data}}\sum_j c_j\ket{j}_{A}\right)\ket{0}_B, \\
        \ket{\Psi_{k1}} &= \frac{1}{2\beta_k} \left(\sum_j c_j\ket{j}_{\text{data}} \ket{k}_A - \ket{k}_{\text{data}}\sum_j c_j\ket{j}_{A}\right)\ket{1}_B,
    \end{align}
    where 
    \begin{align}
        \alpha_k &= \sqrt{\frac{1}{2}(1+r_k^2)}, \\
        \beta_k &= \sqrt{\frac{1}{2}(1-r_k^2)}.
    \end{align}
    We define \(\theta_k \in [1/4,1/2)\) by
    \begin{align}
        \sin \pi\theta_k &= \alpha_k.
    \end{align}
    Then \(\ket{\Psi_k}_{\text{data}, A, B}\) can be rewritten as
    \begin{equation}
        \ket{\Psi_k}_{\text{data}, A, B} = \alpha_k \ket{\Psi_{k0}} + \beta_k \ket{\Psi_{k1}}.
    \end{equation}
    We denote \(Z_B\), \( V S_k V^\dagger\) and \(G_k\) acting on the subspace spanned by \(\{\ket{\Psi_{k0}}, \ket{\Psi_{k1}}\}\) as \(\widetilde{Z_B}\), \( \widetilde{V S_k V^\dagger}\) and \(\widetilde{G_k}\) respectively. The first two can be written as
    \begin{align}
        \widetilde{Z_B} &= \ket{\Psi_{k0}}\bra{\Psi_{k0}} - \ket{\Psi_{k1}}\bra{\Psi_{k1}}, \\
        \widetilde{V S_k V^\dagger} &= (1 - 2\alpha_k^2)\ket{\Psi_{k0}}\bra{\Psi_{k0}} \nonumber \\ 
        &\quad + (1-2\beta_k^2)\ket{\Psi_{k1}}\bra{\Psi_{k1}} \nonumber \\
        &\quad - 2\alpha_k \beta_k(\ket{\Psi_{k1}}\bra{\Psi_{k0}}+\ket{\Psi_{k0}}\bra{\Psi_{k1}}).
    \end{align}
    Therefore \(\widetilde{G_k}\) is
    \begin{align}
        \widetilde{G_k} &= \widetilde{V S_k V^\dagger}\widetilde{Z_B} \nonumber \\
        &= (1 - 2\alpha_k^2)\ket{\Psi_{k0}}\bra{\Psi_{k0}} \nonumber \\ 
        &\quad - (1-2\beta_k^2)\ket{\Psi_{k1}}\bra{\Psi_{k1}} \nonumber \\
        &\quad - 2\alpha_k \beta_k(\ket{\Psi_{k1}}\bra{\Psi_{k0}}-\ket{\Psi_{k0}}\bra{\Psi_{k1}}).
    \end{align}
    Two eigenvalues of \(\widetilde{G_k}\) are
    \begin{equation}
        \lambda_{k\pm} = e^{\pm i2\pi\theta_k},
    \end{equation}
    and eigenvectors \(\ket{\Psi_{k\pm}}\) each corresponding to \(\lambda_{k\pm}\) are
    \begin{equation}
        \ket{\Psi_{k\pm}} = \frac{1}{\sqrt{2}} \left(\ket{\Psi_{k0}} \pm i\ket{\Psi_{k1}}\right).
    \end{equation}
    \(\ket{\Psi_{k}}\) can be decomposed into \(\ket{\Psi_{k\pm}}\) as
    \begin{equation}
        \ket{\Psi_{k}} = \frac{-i}{\sqrt{2}}\left(e^{i\pi\theta_k}\ket{\Psi_{k+}}-e^{-i\pi\theta_k}\ket{\Psi_{k-}}\right).
    \end{equation}

    \bibliographystyle{apsrev4-1}

\end{document}